# Fingerprint of fractional charge transfer at metal/organic interface

Sabine-A. Savu,[1] Giulio Biddau,[2] Lorenzo Pardini,[2] Rafael Bula,[3] Holger F. Bettinger,[3] Claudia Draxl,[2] Thomas Chassé,[2] M. Benedetta Casu[1*]

[1] Institute of Physical and Theoretical Chemistry, University of Tübingen,

Auf der Morgenstelle 18, 72076 Tübingen, Germany

[2] Physics Department and IRIS Adlershof, Humboldt-Universität zu Berlin , Zum Großen Windkanal 6,

12489 Berlin Germany

[3] Institute of Organic Chemistry, University of Tübingen, Auf der Morgenstelle 18,

72076 Tübingen, Gemany

*E-mail: benedetta.casu@uni-tuebingen*

**Abstract**

Although physisorption is a widely occurring mechanism of bonding at the organic/metal interface, contradictory interpretations of this phenomenon are often reported. Photoemission and X-ray absorption spectroscopy investigations of nanorods of a substituted pentacene, 2,3,9,10-tetrafluoropentacene, deposited on gold single crystals reveal to be fundamental to identify the bonding mechanisms. We find fingerprints of a fractional charge transfer from the clean metal substrate to the physisorbed molecules. This phenomenon is unambiguously recognizable by a non-rigid shift of the core-level main lines while the occupied states at the interface stay mostly unperturbed, and the unoccupied states experience pronounced changes. The experimental results are corroborated by first-principles calculations.

The key to tuning device performance is the understanding of the various mechanisms that occur at their interfaces. The interfaces form the device, and the type of these interfaces (e.g. metal/metal, metal/semiconductor, or semiconductor junctions) along with related phenomena (Schottky barrier vs. ohmic contacts) define, together with the stability and the properties of the active layers,[1,2] the electronic characteristics, the performance, and the lifetime of a device. Although organic/metal interfaces have been the focus of nearly two decades of investigations,[3-7] very recently the interest in this type of interface has enjoyed a renaissance with the flourishing of a body of work focused mainly on the interfaces between chemisorbed organic molecules and metals.[8-11] Photoemission features strongly depend on the strength of the molecule/substrate interaction, as demonstrated for a number of molecules on Ag(111).[9] These experiments show that the stronger the bond of the molecules with the substrate, the larger the effect of the charge transfer on photoemission is.[9] Resonance structures are stabilized on the surface through an initial metal-to-molecule charge transfer and rehybridization of suitable side groups, leading to an extended π-electron system that is strongly coupled to the metal states. [8] This coupling, decreasing the molecular electronic gap, overcomes the competing phenomenon of Fermi-level pinning, and leads to substantially charged molecular monolayers, which, if the Fermi level comes to lie within a frontier molecular orbital, behave as metallic. The change from semiconducting to metallic nature of the organic material is suggested as a new route for the chemical engineering of metal surfaces.[8]

The present "state of the art" in the interpretation of the organic/metal and organic/organic interfaces, is based on the interface interaction strength, considering the complete range of interactions from physisorption of noble gases to strong chemisorption of π-conjugated molecules.[11] The case of physisorption with and without charge transfer is typically examined within the integer electron charge transfer model, stating that physisorption on organic and passivated metal surfaces is possible, while weak chemisorption, with possible fractional charge transfer, occurs on non-reactive clean metal surfaces.[11] These models, although extremely useful and detailed, do not explain all experimental results: Physisorption is a widely occurring phenomenon at the organic/metal interface, exhibiting different spectral characteristics that also depend on the strength of molecule/metal interaction. We here contribute to the understanding of the organic/metal interface by an experimental and theoretical multi-technique investigation, choosing 2,3,9,10-tetrafluoropentacene[12] (F4PEN, Figure 1), a fluorinated pentacene derivative, as a model system. Pentacene-based molecules are potential candidates for organic electronics due to the fact that substitution is a very powerful way to tailor the optical and electrical molecular properties to specific technological needs.[13] Indeed, unsubstituted pentacene (PEN), a p-type semiconductor, is the subject of numerous investigations owing to its high charge-carrier mobility and its ability to form highly oriented thin films.[3,4,14-20] These properties can be tuned, according to specific technological needs, for example, by fluorination, turning, e.g., pentacene into the n-type semiconductor perfluoropentacene.[13,21,22]

In this Letter, we investigate F4PEN molecules deposited on Au(110) single crystals by using X-ray photoelectron spectroscopy (XPS), ultraviolet photoelectron spectroscopy (UPS), and near edge X-ray absorption fine structure (NEXAFS) spectroscopy. The simultaneous use of this variety of techniques, in combination with controlled in-situ deposition, gives the opportunity not only to explore the complete electronic structure of the systems (occupied and unoccupied states, or orbitals) but to avoid possible artefacts and discrepancies due to slightly different preparations that could impact the morphology and the structure of the assemblies and consequently their electronic structure.[23-25]

We demonstrate that a fractional charge transfer from the metal substrate to the physisorbed molecules occurs and that it exhibits a very specific and clearly recognizable fingerprint: This is a non-rigid shift of the XPS main lines and at the interface a strong alteration of the NEXAFS signal together with almost unperturbed occupied states. The experimental results are corroborated by calculations based on density-functional theory (DFT), as implemented in the `exciting` code.[26,27]

In Figure 1, the thickness-dependent core level spectra of F4PEN nanorod assemblies are shown (for experimental details, nanorod morphology and XPS stoichiometric analysis after deposition see Supplemental Material[27]). The C1s core level spectra of the thicker assemblies are dominated by a peak at 285.5 eV and a further peak at 286.4 eV. The main peak can be assigned to carbon atoms in the ring of the backbone (carbon atoms bound only to carbon (CC) or also to hydrogen (CH)), while further features at higher binding energy are related to carbon atoms which are bonded to fluorine (CF). With increase of the nominal thickness, the C1s main line is shifted toward higher binding energy by ~1 eV. A widely spread satellite structure which is typical for acenes[28] is also visible at higher binding energy (see Supplemental Material for a zoom into this region[27]). We observe that with increasing thickness the spectroscopic lines do not experience a rigid homogeneous shift: The core level shift is more pronounced for the main line that is 1.04 (main line at lower binding energy, for contributions related to the pentacene backbone) compared to 0.98 eV (for contributions related to CF, see Figure 1a). The thickness-dependent F1s spectra show a single line at 688.0 eV (Figure 1b), as expected because of the presence of fluorine atoms that have the same chemical environment. A 0.85 eV shift toward higher binding energies is visible comparing the thickest and the thinnest assemblies.

Apart from these aspects, no other relevant changes in XPS line shape and intensity are observable. Thus, we conclude that the molecular density of states stay unperturbed at the interface. The shake-up satellite features are also visible in the spectra of the thinnest assemblies at higher binding energies; however, their intensity does not show abrupt changes at the interface (see Supplemental Material[27]). These three observations hint at weak physisorption, which is further supported by the fact that the molecules are almost completely desorbed after a short annealing at 415 K (see Supplemental Material[27]). Therefore, a possible chemical bond at the F4PEN/Au interface can be ruled out as a reason for the observed non-rigid shift of the spectroscopic lines.

To investigate the origin of this non-rigid shift, we perform UPS measurements. The He I UPS spectra (Figure 1c) do not show any evidence for band gap states.[7] This observation supports the conclusion that the molecules are physisorbed on the surface. In fact, chemisorption would significantly modify the photoemission features of the F4PEN molecular assemblies close to the interface,[8,9] as also discussed for XPS above. We observe only a slight shift (~0.16 eV) of the highest occupied molecular orbital (HOMO) onset towards higher binding energies (Figure 1d). This observation does not exclude the possibility that the non-rigid shift may originate from a local image-charge at the interface that gives rise to a different screening of the various atoms in the molecule. This aspect can be explored by NEXAFS spectroscopy because image-charge screening at the interface does not affect the intensity of the NEXAFS resonances.[29,30] Thus, if only image-charge screening occurs, we do expect no significant changes in the NEXAFS spectra collected for the thin and the thick assemblies.

Surprisingly, we observe a clear thickness-dependent difference in the intensity of the $\pi^*$-resonances, which represent the prominent spectral features in the range up to around 288 eV photon energy in the NEXAFS spectra (Figure 2, see Supplemental Material for polarization dependent NEXAFS experiments[27]). These resonances have lower intensity for the thin assemblies. A similar behaviour was found for nanorod assemblies of other substituted pentacenes.[31,32] Likewise, a non-rigid shift is reported in the literature for other physisorbed organic molecules like 3,4,9,10-perylene-tetracarboxylic acid dianhydride (PTCDA),[33] cobalt phtalocyanine,[34] and magnesium phthalocyanine.[35] This behaviour is present irrespective of the specific orientation of the molecules on the substrate that ranges from flat lying (PTCDA[33] and phtalocyanine[34,35]) to recumbent (substituted pentacenes[31,32]). Note that polarization dependent NEXAFS experiments show that that first layer of F4PEN molecules at the interface with gold are flat-lying (see Electronic Supporting Information). The non-rigid shift is also observed irrespective of the character of the substituents (electron-accepting or electron-donating groups), for example, it is also recognisable in 2,3,9,10-tetramethoxy-pentacene assemblies. [32]

While photoemission experiments unambiguously demonstrate that the molecules are physisorbed, i.e., there is no evidence of a change in the molecular density of states at the interface, in NEXAFS, we observe a clear decrease in intensity of the $\pi^*$-resonances, although the valence band of the molecular assembly is unperturbed.[32] We interpret this behaviour as being due to charge transfer from the metal substrate to the molecules.

DFT calculations support our interpretation. To mimic the thin assembly that experiences charge transfer from the substrate, we consider a single F4PEN molecule and add different amounts of electronic charge. The sampling depth of our experiment is around 54 Å, corresponding to the estimated inelastic mean free path of 18 Å[36]. It means that the XPS spectra of the 254 Å nominally thick assembly do not contain contributions from the interface. Consequently, the thick assembly results can be compared to neutral molecules since in this case the charge transfer at the interface is masked. The theoretical approach of considering single molecules is justified by the fact that, as seen by XPS and UPS, the metal/molecule interaction is weak and the molecular orbitals do not change their character upon adsorption[37-39] as we will discuss later in more detail. The results are presented in Figure 3 (Upper panel), where the calculated core-level shifts are depicted as a function of transferred charge. A non-rigid shift of the core levels towards lower binding energies is clearly observed. By a direct comparison of the theoretical results with the experimental ones, we can deduce that the experimental values are well reproduced when 0.75 electrons are added to the molecule. In this case, in fact, we find shifts of 1.16 and 1.14 eV (comparable with the experimental shift of 1.04 eV of the XPS main line) for contributions related to CH and CC; 1.01 eV for CF (to be compared with the experimental value of 0.98 eV); and 0.79 eV for F atoms (corresponding to 0.85 eV in the experiment). In other words, all the theoretical values nicely reproduce the experimentally observed core level shifts (see also Figure 1). Our calculations show that in the "charged" system, the increased screening gives rise to diminished nuclear attraction, pulling the core levels upwards. Note that our findings do not change when a core hole is included in the calculations.[40] Figure 3 (Middle panel) shows the comparison of the XPS core level spectra with a convolution of delta functions representing the computed core states of the charged and neutral molecule, respectively. This convolution is based on a Voigt profile adopting a Lorentzian with a full-width at half maximum ($W_L$) of 0.1 eV and a Gaussian with a full-width at half maximum ($W_G$) of 1 eV. The Voigt profile is chosen in order to take into account both the finite core-hole lifetime (which has a Lorentzian profile) and the broadening due to the finite experimental resolution as well as various inhomogeneities, e.g., molecular packing and local morphology[24] (Gaussian profile).

Since the interpretation of the experimental results in terms of single molecules might appear simplistic, we perform analogous calculations for an F4PEN monolayer of flat-lying molecules for the case of 0.75 electrons charge transfer. In fact, the results for the single molecules are very well reproduced as indicated by the stars in Fig. 3. The maximum deviation, i.e. 0.08 eV, is obtained for CF, while both CC and CH increase by only 0.01 eV, and F1s experiences a slightly larger shift of 0.07 eV.

We consider 0.75 electrons as the upper limit of the transferred charge since in this quantification we accumulate also contributions due to other possible sources of (rigid) core-level shifts: These are changes in the molecular orientation,[41] different charge redistribution, due to the strong electronegativity of the fluorine atoms[31] when comparing surface and bulk environment of the molecules (surface core level shift[36,42]) and image-charge screening effects due to the capability of the substrate to screen the core-hole generated in the photoemission event.[7]

The experimental finding that no appreciable change in the HOMO takes place upon charge transfer (compare the UPS spectra of the thick and thin assemblies in Figure 1) is also supported by our calculations. To this extent, Figure 3 (lower panel) shows the calculated molecular orbital images in the Tersoff-Hamann approximation for the HOMO of the neutral and charged molecule, respectively. Concomitant to the UPS experiment, the orbital does not change its shape.

From the combination of various experimental probes and first-principles calculations to determine the electronic structure of physisorbed nanorod assemblies of F4PEN, we find evidence for a charge transfer from a non-reactive clean metal surface to the unoccupied states of the physisorbed molecules. This has specific photoemission and X-ray absorption spectral fingerprints. We quantify the observed core-level shifts as being due a fractional charge transfer.

Our work, revealing the true nature of the occurring interface phenomenon, contributes with a consistent understanding to the picture of the organic/metal interface previously drawn by other works.[8,9,11] Our results also show that for weakly bound systems, a single molecule approximation can be a reliable approach for the description of the occupied states of a physisorbed molecule, since these orbitals do not change upon adsorption. This is evidenced experimentally and theoretically.

What is most important, a variety of different complex phenomena occur at the metal interface in physisorbed systems. These may involve either charge transfer and/or charge image screening. These mechanisms affect the occupied and the unoccupied states in different ways: Their common characteristic is to leave the occupied states almost unperturbed. To explicitly identify their nature (e.g., charge transfer versus image screening) the investigation of the unoccupied states plays a fundamental role. In this respect, the combination of photoemission and X-ray absorption spectroscopy, supported by first-principles calculations, reveals the fingerprints to unambiguously describe the adsorption mechanisms at the metal/organic interface.

The authors thank the Helmholtz-Zentrum Berlin (HZB), Electron storage ring BESSY II, for providing beamtime, the HZB resident staff for beamtime support, S. Pohl, W. Neu and E. Nadler for technical support. Financial support from the Helmholtz-Zentrum Berlin is gratefully acknowledged. GB, LP, and CD appreciate support from the Austrian Science Fund (Project I543) and the German Research Foundation (DFG, through the Collaborative Research Project 951).

Figure captions

Figure 1. (color online) a) Thickness dependent C1s, and b) F1s core level spectra of F4PEN assemblies. The core-level shifts toward higher binding energy with increasing thickness are indicated. c) Zoom into the near $E_F$ region of the valence band measured using He I ($E_F = 0$ eV) (solid line: clean Au(110)). d) HOMO onset binding energy plotted against the nominal thickness. The work function is 5.1 eV for the clean substrate and it decreases by 0.7 eV upon F4PEN deposition. The molecular structure is also shown.

Figure 2. (color online) C K-NEXAFS spectra for nominal thicknesses of 20 Å (thin assembly, lower) and 175 Å (thick assembly, upper), measured in in-plane polarization. The arrows evidence the resonances that experience the strongest changes, as discussed in the text.

Figure 3. (color online) Upper panel. Theoretical core level shifts of carbons and fluorine atoms versus the added amount of charge. Positive values represent shifts towards lower binding energies. Positive values on the *x* axis mean that a fractional (negative) electron charge is added to the cell. The arrow indicates the amount of charge transfer comparable with the experimental findings.
Middle panel. a) XPS C1s and b) XPS F1s core level spectra for the thin (2 Å) and thick (254 Å) assembly, as indicated. The blue curves are convolutions of the computed core levels aligned to experiment with respect to the main line.
Lower panel: Images for the HOMO orbital of the neutral (left) and charged (right) molecule. They are calculated from the LDOS integrated in the energy range [-1.0, 0.0] (neutral molecule) and [-1.5, 0.0] eV (charged molecule).

Figure 1

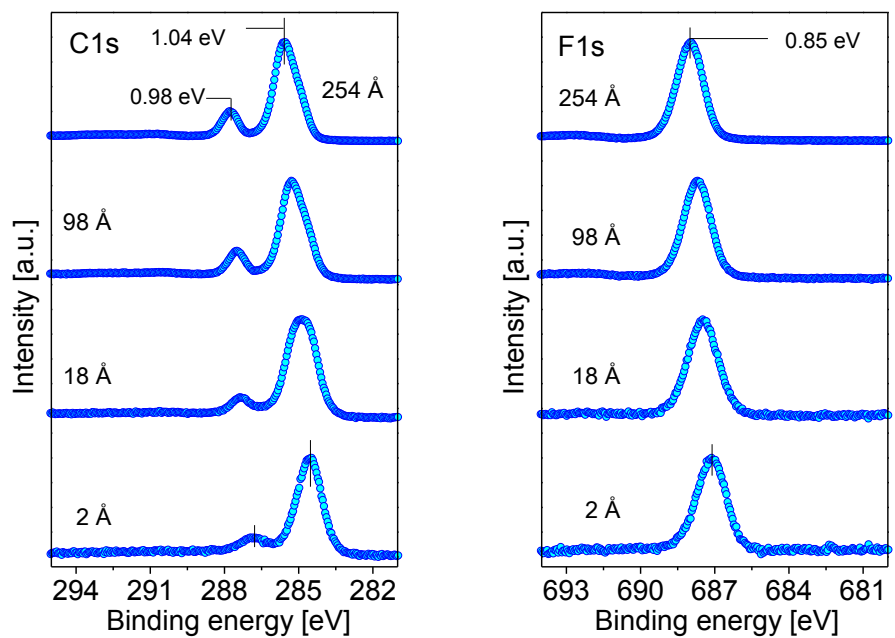

a).........................................................b)

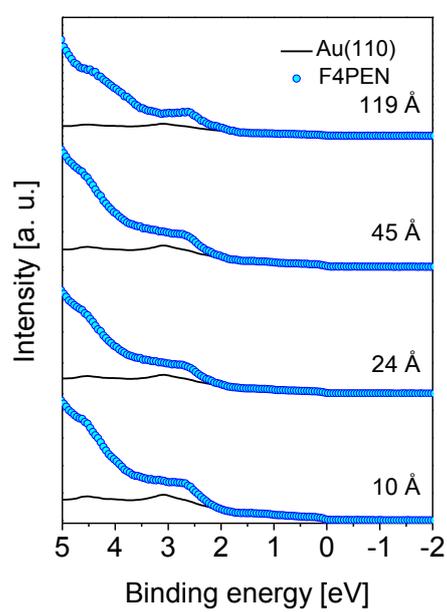

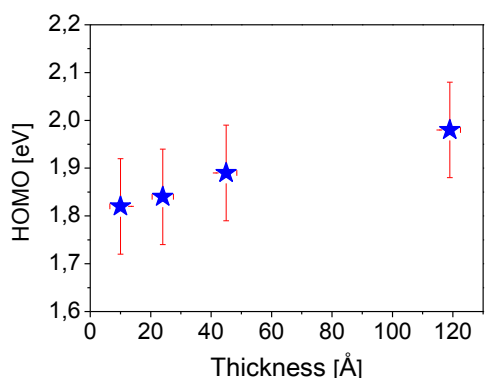

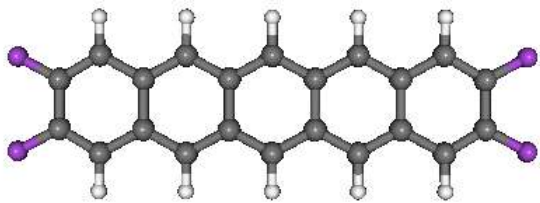

d)

c)

Figure 2

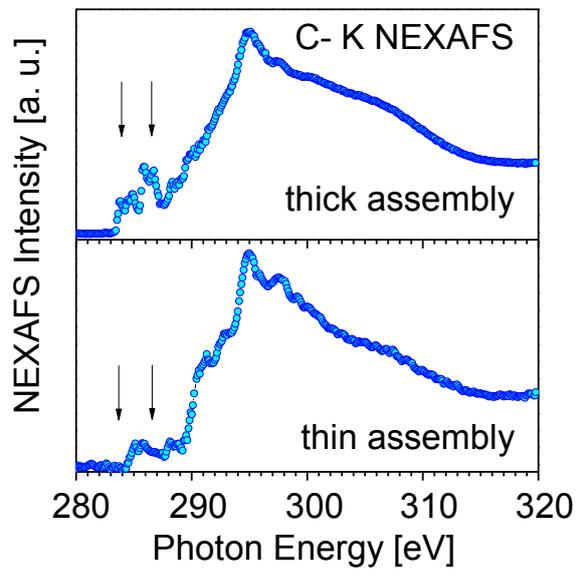

Figure 3

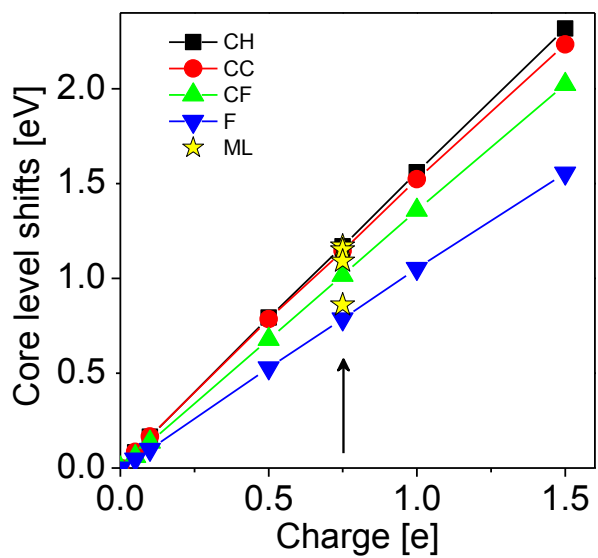

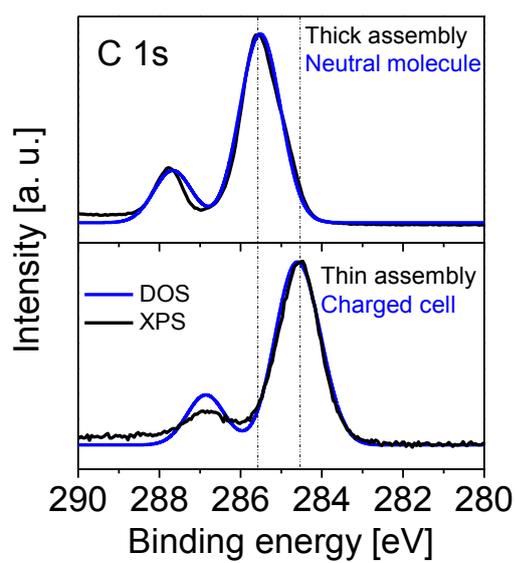
a)

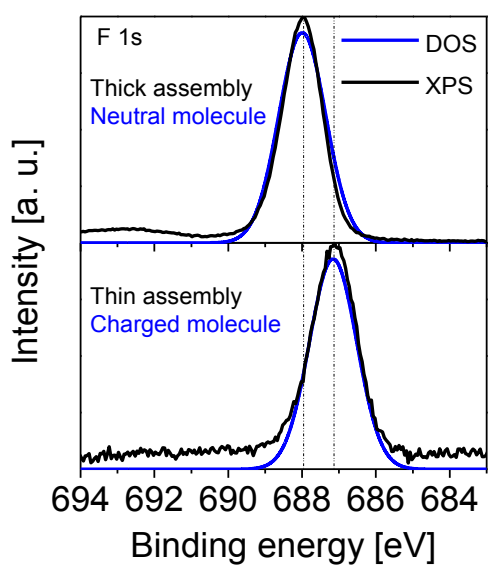
b)

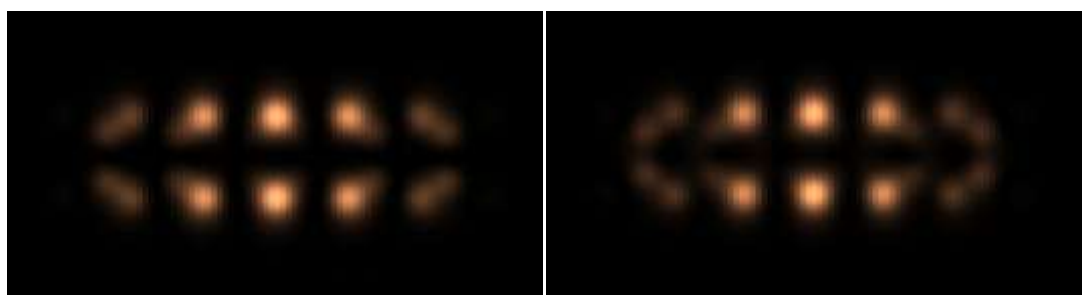